\documentclass[letterpaper, 10 pt, conference]{ieeeconf}  
\IEEEoverridecommandlockouts                              

\usepackage{cite}
\usepackage{amsmath,amssymb,amsfonts,mathptmx}
 
\usepackage{amsthm}
\usepackage{graphicx}
\usepackage{enumerate}
\usepackage{epsfig}
\usepackage[font=small,labelfont=bf]{caption}
\usepackage{amsmath} 
\usepackage{bm}
\usepackage{epstopdf}
\usepackage{algorithm,array}
\usepackage[noend]{algpseudocode}
\usepackage{multirow,multicol}
\let\labelindent\relax
\usepackage{subfig,braket,enumitem,adjustbox,booktabs}
\usepackage{color}
\usepackage{float}
\newfloat{algorithm}{t}{lop}
\usepackage{pgf,tikz}
\usetikzlibrary{arrows}

\newcommand{\oprocendsymbol}{\hbox{$\square$}}
\newcommand{\oprocend}{\relax\ifmmode\else\unskip\hfill\fi\oprocendsymbol}

\newtheorem{theorem}{Theorem}[section]

\newtheorem{lemma}[theorem]{Lemma}

\newcommand{\real}{\ensuremath{\mathbb{R}}}

\newcommand{\Hc}{\mathcal{H}}

\newcommand{\Tc}{\mathcal{T}}

\newcommand{\diag}{{\rm diag}}

\newcommand\tr{\text{tr}}

\renewcommand\det{\text{det}}

\newcommand\Acbb{\overline{\overline{A}}_c}
\newcommand\Ab{\overline{A}}
\newcommand\Bb{\overline{B}}
\newcommand\Cb{\overline{C}}
\newcommand\Pb{\overline{P}}
\newcommand\Lb{\overline{L}}
\newcommand\Kb{\overline{K}}

\newcommand\Acb{\overline{A}_c}
\newcommand\fb{\overline{f}}
\newcommand\alphab{\overline{\alpha}}
\newcommand\betab{\overline{\beta}}
\newcommand\etab{\overline{\eta}}
\newcommand\deltab{\overline{\delta}}

\newcommand{\longthmtitle}[1]{\mbox{}{\textit{(#1).}}}

\graphicspath{{epsfiles/}}

\title{\LARGE \bf Co-Design of Lipschitz Nonlinear Systems}

\author{Prasad Vilas Chanekar$^{*}$  \quad  Nikhil Chopra
	\thanks{$^{*}$ Corresponding author.}  
	\thanks{PVC is with the Department of Mechanical and Aerospace Engineering, University of California, San Diego, La Jolla, CA 92093, USA, {\tt pchanekar@ucsd.edu}}%
	\thanks{NC is with the Department of Mechanical Engineering, University of Maryland, College Park, MD 20742, USA, {\tt nchopra@umd.edu}}
}

\begin{document}
	\maketitle

	\noindent\begin{abstract} Empirical experiences have shown that simultaneous (rather than conventional sequential) plant and controller design  procedure   leads to an improvement in performance and saving of plant resources. Such a simultaneous synthesis procedure is called as ``co-design". In this letter we study the co-design problem for a class of Lipschitz nonlinear dynamical systems having a quadratic control objective and state-feedback controller. We propose a novel time independent reformulation of the co-design optimization problem whose constraints ensure stability of the system. We also present a gradient-based co-design solution procedure which involves system coordinate transformation and whose output is provably stable solution for the original system. We show the efficacy of the solution procedure through  co-design of a single-link robot. 
		
	\end{abstract}
	
	\section{Introduction}\label{introduction-1}
	Plant design and control parameters are interrelated through the plant dynamics. Sequentially first optimizing the plant design parameters and then the control parameters might result in wastage of (plant) resources through over-design  and also compromise the plant's control  performance~\cite{skelton1989model,fathy2001coupling}. Hence \textit{co-design} (simultaneous optimization) of plant and control parameters may lead to a better plant design (saving of plant resources) and better control performance. 
	
	Co-design finds its application in optimizing  aerospace structures~\cite{hale1985optimal}, electric motors~\cite{reyer2002combined}, robots~\cite{ravichandran2006simultaneous}, chemical processes~\cite{sandoval2008simultaneous} etc. We can classify   system performance improvement  problems with different types of design variables as co-design problems for example: network actuator/sensor placement~\cite{chanekar2017optimal}, structured/sparse controller design~\cite{chanekar2017optimaloutput,lin2013design}, network edge modification~\cite{chanekar2020energy} etc., to name  a few. A detailed literature survey regarding co-design of linear systems can be read in~\cite{chanekar2018co,liu2020decentralized}. 
	
	Co-design optimization is a popular topic in optimization of systems with linear dynamics than its nonlinear counterpart due to a relatively less complex problem setup. 
	Recent works~\cite{wang2014co,jiang2015optimal} have used modified policy iteration scheme to co-design nonlinear systems. The challenges in a co-design optimization problem are: time-dependent objective function and nonlinear dynamic constraint, abstract stability constraint, non-convex and nonlinear nature. The contribution of our work in this letter is as follows.
	\begin{enumerate}
		\item We formulate the co-design optimization problem  for a class of nonlinear  dynamical systems with Lipschitz nonlinearity in state variables.   The problem has a time dependent quadratic control objective and the system is controlled by a full state-feedback controller. We propose a novel time-independent reformulation of the control objective and abstract stability constraint using a quadratic matrix equation.
		\item Next we derive a sufficient condition for the quadratic matrix equation to have  a solution. We then propose a gradient based solution method to solve the co-design optimization problem which involves a coordinate transformation. Finally, we rigorously prove that the co-design solution obtained for the transformed system also stabilizes the original system. 
	\end{enumerate}
	We organize  the remaining letter as follows: We propose the co-design problem in Section \ref{codesign-problem-1} followed by the co-design algorithm and its preliminaries in Section \ref{design-algo-1}. We present an example in Section \ref{example-1} and concluding remarks in Section \ref{conclusion-1}. 
	
	\textit{Notation:} We use $\real$ for the set of real numbers. For a matrix $X\in \real^{m\times n}$ having $m-$rows and $n-$columns, we use $X^{\top}$ for its transpose, $\Vert X\Vert$ for its Frobenius norm and $\Vert X\Vert_2$ for its $2-$norm. For $X\in \real^{m\times m}$, $X(\succeq)\succ0$ denotes $X$ is positive (semi-)definite.  We use $I$ to denote identity matrix of appropriate dimension.
	\section{The Co-Design Problem}
	\label{codesign-problem-1}
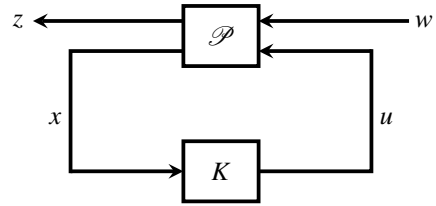
\begin{figure}[htb]
		\centering
		{\begin{tikzpicture}[>=stealth,line width=0.43mm]
				\draw [line width = 0.5mm] (4,0) rectangle (5,0.8);
				\node[] (i1) at (4.5,0.4){$\mathcal{P}$};
				\draw[->,line width = 0.5mm](4,0.6) -- (2,0.6);
				\draw[line width = 0.5mm] (4,0.2) -- (2.5,0.2) -- (2.5,-1.425);
				\draw[->,line width = 0.5mm] (2.5,-1.4) -- (4,-1.4);
				\draw[line width = 0.5mm]  (5,-1.4) -- (6.5,-1.4) -- (6.5,0.225);
				\draw[->,line width = 0.5mm] (6.5,0.2) -- (5,0.2);
				
				\draw [line width = 0.5mm] (4,-1) rectangle (5,-1.8);
				\node[] (i2) at (4.5,-1.4){$K$};
				
				\draw[->,line width = 0.5mm](7,0.6) -- (5,0.6);
				\node[] (i3) at (1.8,0.6){$z$};
				\node[] (i4) at (7.2,0.6){$w$};
				\node[] (i5) at (2.3,-0.7){$x$};
				\node[] (i5) at (6.7,-0.7){$u$};
		\end{tikzpicture}}\hfill
		\caption{Plant $\mathcal{P}$ and controller $K$}
		\label{plant-1}
	\end{figure}

	Consider a plant $\mathcal{P}$  as shown in Fig. \ref{plant-1}.
	\\
	The plant $\mathcal{P}$ follows the Lipschitz nonlinear dynamics
	\begin{align}\label{eq-2.0.1}
		\notag&\dot{x}=Ax+Bu+\Phi\left(x\right)+B_ww,\\
		& z = Cx+Du, \quad u =Kx.
	\end{align}
	Here $x\in \real^{n_x},u\in\real^{n_u},z\in\real^{n_z},w\in\real^{n_w}$  are state, control input, monitored output and disturbance vectors  respectively. $A,B,B_w,C,D$ are system matrices of appropriate dimensions. The elements of the system matrices $A,B$ can be functions of the design variable $d\in \real^{n_d}$. The vectors $x,u,z$ are functions of time $t\in\real,t\geq0$. $K$ is a stabilizing state feedback control gain matrix of appropriate dimension. Note that  by stabilizing controller $K$ we mean that the closed loop system $A_c=A+BK$ is Hurwitz and when $w=0$ then $x\rightarrow 0$  as $t\rightarrow \infty$.  The function $\Phi\left(x\right):\real^{n_x}\mapsto \real^{n_x}$ is a Lipschitz continuous function i.e., there exists a constant $\alpha\in\real,\alpha>0$ such that,
	\begin{align}
		\label{eq-2.0.2}
		\Vert \Phi\left(x_2\right)-\Phi\left(x_1\right)\Vert \leq \alpha \Vert x_2-x_1\Vert \quad \forall x_1,x_2 \in \real^{n_x}.
	\end{align}
	We quantify the control performance of the system  using a monitored output $z$ and express it  mathematically using the $\mathcal{L}_2$-norm \cite{skogestad2005multivariable}  as follows,
	\begin{align*}
		f_c = \int_{0}^{\infty}z^{\top}z\;dt.
	\end{align*}
	The objective $f_c$ is a function of the controller variable $K$ through the input and the design variable $d$ through the system dynamics. In the co-design process we optimize $d$ and $K$ simultaneously. We formulate the co-design problem  as follows, 
	\begin{align}\label{eq-2.0.5}
		\underset{d,K}{\min} \qquad &
		\notag\beta_df_d\left(d\right)+ \beta_c\int_{0}^{\infty}z^{\top}z\;dt
		\\
		\notag \text{s.t.}  \qquad
		\notag&\dot{x}=Ax+Bu+\Phi\left(x\right)+B_ww,\\
		& z = Cx+Du, \quad u =Kx,
		\\
		\notag &\underline{d}\leq d\leq \overline{d},\\
		\notag & A+BK\;  \text{ is Hurwitz},\\
		\notag &  x\rightarrow 0 \; \text{as}\; t\rightarrow \infty.
	\end{align}
	Here $f_d:\real^{n_d}\mapsto\real$  is a design function, $\beta_d\in\real,\beta_d\geq 0,\beta_c\in\real,\beta_c\geq 0$ are weighing constants and $\underline{d}\in\real^{n_d},\overline{d}\in\real^{n_d}$ are lower and upper bounds respectively on the  design variable.

	We have   following assumptions on the system \eqref{eq-2.0.1} and  problem \eqref{eq-2.0.5},
	\begin{itemize}
		\item[(\textbf{A1})] $\left(A,B\right),\;\left(A,B_w\right)$ is \textit{structurally  stabilizable} in the design variable space.
		\item[(\textbf{A2})] $\left(A,C\right)$ are detectable.
		\item[(\textbf{A3})] $D^{\top}\begin{pmatrix}
			D & C
		\end{pmatrix}=\begin{pmatrix}
			R & 0
		\end{pmatrix},\quad R=R^{\top}, \quad R \succ 0$.
		\item[(\textbf{A4})] $\Phi\left(0\right)=0.$
		\item[(\textbf{A5})] The function $f_d\left(d\right)$ is smooth and bounded.
	\end{itemize}
	Here by structural stability we mean that there exists some $d$ in the design space such that $A_c$ is Hurwitz for some $K$.
	The problem \eqref{eq-2.0.5}  is a time dependent, non-convex and nonlinear optimization problem with an abstract Hurwitz constraint.  In the current format, the co-design problem is challenging and difficult to solve.   Hence to obtain a solution, we first reformulate the time dependent co-design problem into a  time-independent problem free from the abstract constraint.
	\subsection{Reformulated Co-design Problem} 
	\label{pcopr-1}
	To reformulate  \eqref{eq-2.0.5}, we  first  derive a mathematical condition to replace the abstract Hurwitz constraint using Lyapunov stability theory \cite{skogestad2005multivariable}. 
	\begin{theorem}\longthmtitle{Reformulation of Hurwitz constraint}
		\label{thm1}
		Consider the system \eqref{eq-2.0.1} and with a   controller gain $K$. Then for $w=0$  the system is stable if  there exists a  matrix $P=P^{\top}\succ 0$ such that with $C_z = C+DK$ and $C_z^{\top}C_z\succ 0$,
		\begin{align}\label{eq-thm1-1}
			A_c^{\top}P+PA_c+\alpha^2PP+I+C_z^{\top}C_z=0.
		\end{align}
	\end{theorem}
	\begin{proof}
		Consider the candidate Lyapunov function,
		\begin{align}
			\label{eq-thm1-2}
			V = x^{\top}Px, \quad P=P^{\top}\succ 0.
		\end{align}
		The time derivative of $V$ denoted by $\dot{V}$  and using \eqref{eq-2.0.1}  is,
		\begin{align*}
			\notag\dot{V}&=\dot{x}^{\top}Px+x^{\top}P\dot{x},\\
			&= x^T\left({A_c}^{\top}P+PA_c\right)x+{\Phi\left(x\right)}^{\top}Px+x^{\top}P\Phi\left(x\right).
		\end{align*} 
		Using ${\Phi\left(x\right)}^{\top}Px=x^{\top}P\Phi\left(x\right)$, $\Vert \Phi\left(x\right)\Vert \leq \alpha \Vert x \Vert$, 
		\begin{align*}
			\dot{V}=x^T\left({A_c}^{\top}P+PA_c\right)x+2{\Phi\left(x\right)}^{\top}Px.
		\end{align*} 
		Using ${\Phi\left(x\right)}^{\top}Px\leq\Vert Px\Vert \Vert \Phi\left(x\right)\Vert \leq \alpha\Vert Px\Vert \Vert x\Vert $,  $2\alpha\Vert Px\Vert\Vert x\Vert\leq \alpha^2x^{\top}PPx+x^{\top}x$, $u=Kx$, adding and subtracting $z^{\top}z$,
		\begin{align}
			\notag\dot{V}&\leq x^T\left({A_c}^{\top}P+PA_c+\alpha^2PP+I+C_z^{\top}C_z\right)x-z^{\top}z.
		\end{align} 
		When \eqref{eq-thm1-1} holds $\dot{V}\leq -z^{\top}z<0$ implying $K$ is a stabilizing controller \cite{Khalil:1173048}.	
	\end{proof}
	We now replace the time dependent objective in \eqref{eq-2.0.5}  by an  time-independent function. 
	\begin{theorem}\longthmtitle{Upper bound on control objective function}
		\label{thm2}
		Consider system \eqref{eq-2.0.1} with the stabilizing controller gain $K$ and let $ x(0) = 0$ then,
		\begin{align}
			\label{eq-thm2-1}
			\int_{0}^{\infty}z^{\top}z\;dt \leq \tr\left(B_w^{\top}PB_w\right),
		\end{align}
		where $P$ is the solution of \eqref{eq-thm1-1}.	
	\end{theorem}
	\begin{proof}
		The system starts from  $x=0$ at $t=0$.  The disturbance is like an  impulse input $\delta\left(t\right)$  acting at $t=0$. Let $t=0^+$ is a time instant immediately  after the impulse input at $t=0$. This results in $x\left(0^+\right)\neq 0$. Thus,  
		\begin{subequations}
			\label{eq-thm2-2}
			\begin{align}
				\label{eq-thm2-2a}&x\left(t\right)=x\left(0\right)=0\quad \forall t\in \left[0,0^+\right),\\
				\notag &\text{which also implies,}\\
				\label{eq-thm2-2b}&u\left(t\right)=0, \quad \Phi\left(x\right)=0 \quad \forall t\in \left[0,0^+\right).
			\end{align} 
		\end{subequations}
		The integral in \eqref{eq-thm2-1} is decomposed as,
		\begin{align*}
			\int_{0}^{\infty}z^{\top}z\;dt = \int_{0}^{0^+}z^{\top}z\;dt+\int_{0^+}^{\infty}z^{\top}z\;dt.
		\end{align*}
		Due to \eqref{eq-thm2-2}, $\int_{0}^{0^+}z^{\top}z\;dt=0$ and from Theorem \ref{thm1},
		\begin{align*}
			\notag\int_{0}^{\infty}z^{\top}z\;dt &= \int_{0^+}^{\infty}z^{\top}z\;dt\leq-\int_{0^+}^{\infty}\dot{V}dt,\\
			&\leq V\left(0^+\right)-V\left(\infty\right).
		\end{align*}
		As $u=Kx$ is a stabilizing control of the system \eqref{eq-2.0.1} so $x\left(\infty\right)=0$. Using \eqref{eq-thm1-2} from Theorem \ref{thm1},
		\begin{align*}
			\notag\int_{0}^{\infty}z^{\top}z\;dt &\leq{x\left(0^+\right)}^{\top}Px\left(0^+\right)-{x\left(\infty\right)}^{\top}Px\left(\infty\right),\\
			&\leq {x\left(0^+\right)}^{\top}Px\left(0^+\right).
		\end{align*}
		Next task is to compute $x\left(0^+\right)$. Let $\{e_1,\ldots, e_k,\ldots, e_{n_w}\}$ be the basis of the disturbance input space where $e_k$ is the $k^{th}$ canonical unit vector.
		The disturbance input is $w=e_k\;\delta\left(t\right)$.
		Using $\int_{0}^{0^+}\delta\left(t\right)\;dt=1$  and $x\left(0^+\right) = \int_{0}^{0^+}B_ww\;dt$,
		\begin{align*}
			x\left(0^+\right) =  \int_{0}^{0^+}B_we_k\;\delta\left(t\right)\;dt=B_we_k=x_k\left(0^+\right).
		\end{align*}
		Let $z_k$ be the output of the system due to an impulse applied in the direction $e_k$ at $t=0$ with other directions receiving no input. Then $\int_{0^+}^{\infty}z_k^{\top}z_k\;dt\leq{x_k\left(0^+\right)}^{\top}Px_k\left(0^+\right)$ and, 
		\begin{align*}
			\notag\int_{0}^{\infty}z^{\top}z\;dt &=\sum_{k=1}^{n_w}\int_{0^+}^{\infty}z_k^{\top}z_k\;dt\leq\sum_{k=1}^{n_w}{x_k\left(0^+\right)}^{\top}Px_k\left(0^+\right),\\
			&\leq \sum_{k=1}^{n_w}\tr\left(e_ke_k^{\top}B_w^{\top}PB_w\right)=\tr\left(B_w^{\top}PB_w\right).
		\end{align*}
	\end{proof}
	Note that unlike in  the case of linear systems \cite{stoorvogel1993robust}, \eqref{eq-thm2-1} has an inequality  due to the presence of Lispchitz non-linearity in the dynamics \eqref{eq-2.0.1}.
	
	Thus the  control objective $\int_{0}^{\infty}z^{\top}z\;dt$ in \eqref{eq-2.0.5} is  upper-bounded by $\tr\left(B_w^{\top}PB_w\right)$. Using   \eqref{eq-thm1-1}, we reformulate \eqref{eq-2.0.5} as,
	\begin{align}\label{eq-2.1.2}
		\underset{d,K,P}{\min} \qquad &
		\beta_df_d\left(d\right)+\beta_c\tr\left(B_w^{\top}PB_w\right)
		\\
		\notag \text{s.t.}  \qquad
		& A_c^{\top}P+PA_c+\alpha^2PP+I+C_z^{\top}C_z=0,\\
		\notag&\underline{d}\leq d\leq \overline{d},\quad P\succ 0.
	\end{align}
	The problem \eqref{eq-2.1.2} is a time-independent, non-convex and nonlinear optimization problem. In the next section we describe a gradient descent method to compute a solution to \eqref{eq-2.1.2}.
	
	\section{Co-Design Algorithm}
	\label{design-algo-1}
	In this section, we propose a gradient descent based iterative  algorithm  to solve the co-design problem \eqref{eq-2.1.2}. We first derive a matrix quadratic equation for synthesizing the initial stabilizing controller for the iterative scheme. We also give a sufficient condition for the matrix quadratic equation to have a solution. Next we propose a coordinate transformation which will be applied on the original system when the aforementioned sufficient is not fulfilled. We then derive expressions for the gradient of the objective function required for the iterative  scheme. We end this section by stating the iterative co-design algorithm. 
	\subsection{Computation of Initial Stabilizing Controller Gain}
	\label{initial-gain-1}
	Unlike linear systems, a Hurwitz closed loop system  does not ensure  stability of a Lipschitz nonlinear system \eqref{eq-2.0.1}.  Hence our initial stabilizing controller $K^0$  consists of two parts \cite{pagilla2004controller}, $K_p^0$ which makes $A+BK_p^0$ Hurwitz and $K_s^0$ determined from  a matrix quadratic equation derived using the Lyapunov stability theory. 
	Before presenting the main result we define the following number $\delta_0(M,N)$   \cite{aboky2002observers,pagilla2004controller},
	\begin{align}
		\label{eq-3.1.2}
		\delta_0(M,N)=\underset{\omega\in\real}{\min}\quad \sigma_{\min}\begin{pmatrix}
			\imath\omega I-M\\
			N
		\end{pmatrix}.
	\end{align}
	Now we present our main result.
	\begin{theorem}\longthmtitle{Initial stabilizing controller gain}
		\label{thm3}
		Consider the system \eqref{eq-2.0.1} with a controller gain
		\begin{align}
			\label{eq-thm3.1}
			K^0 = K_p^0+\alpha^2\frac{K^0_s}{\Vert B\Vert^2},
		\end{align}
		where $K_p^0$ is any known  matrix such that $A_c^0=A+BK_p^0$ is Hurwitz. Then a sufficient condition for a given $\eta\in \real,\eta>0$, $\delta_0\left(A_c^0,\alpha
		\sqrt{1+\eta}\frac{B^{\top}}{\Vert B\Vert}\right)=\delta_0$, $K_s^0=\frac{-\alpha^2B^{\top}P^0}{2}$ with matrix $P^0={P^0}^{\top}\succ 0$ such that 
		\begin{align}
			\label{eq-thm3.5}
			&{A_c^0}^{\top}P^0+P^0A_c+\alpha^2P^0\left(I-\frac{BB^{\top}}{\Vert B\Vert^2}\right)P^0+I+\eta I=0.
		\end{align}
		to hold is 
		\begin{align}
			\label{eq-thm3.5a}
			\delta_0>\alpha\sqrt{1+\eta}.
		\end{align}
	\end{theorem}
	\begin{proof}
		Consider a Lyapunov function $V^0=x^{\top}P^0x$ with the matrix $P^0={P^0}^{\top}\succ 0$. Differentiating $V^0$ with respect to time, using the closed loop dynamics with controller gain $K^0$ and following procedure similar to the proof of Theorem~\ref{thm1} we  have,
		\begin{align*}
			\notag	\dot{V^0}&\leq x^T\left({A_c^0}^{\top}P^0+P^0A_c+\alpha^2P^0P^0\right.\\
			&\hspace{2cm}\left.+I+\frac{{K^0_s}^{\top}}{\Vert B\Vert^2}B^{\top}P^0+P^0B\frac{K^0_s}{\Vert B\Vert^2}\right)x.
		\end{align*} 
		For stability we desire $V^0\rightarrow 0$ as $t\rightarrow \infty$ which is possible when $\dot{V^0}<0$. Now when $K_s^0=\frac{-\alpha^2B^{\top}P^0}{2}$ and for some given $\eta\in \real,\eta>0$ we have,
		\begin{align}
			\label{eq-thm3.4}
			\notag	\dot{V^0}&\leq x^T\left({A_c^0}^{\top}P^0+P^0A_c+\alpha^2P^0\left(I-\frac{BB^{\top}}{\Vert B\Vert^2}\right)P^0\right.\\
			&\hspace{4cm}\left.+I+\eta I\right)x-\eta x^{\top}x.
		\end{align}
		\eqref{eq-thm3.4} will result in   $\dot{V^0}\leq-\eta x^{\top}x<0$ i.e., $\dot{V^0}<0$ if \eqref{eq-thm3.5} holds.
		As $A_c^0$ is Hurwitz, $\Big(I+\frac{BB^{\top}}{\Vert B\Vert^2}\Big)\succ 0$, then  \eqref{eq-thm3.5}  will hold true when the Hamiltonian  of \eqref{eq-thm3.5}
		\begin{align*}
			\Hc^0 = \begin{pmatrix}
				A_c^0 & \alpha^2\Big(I-\frac{BB^{\top}}{\Vert B\Vert^2}\Big)\\
				-(1+\eta) I& -{A_c^0}^{\top}
			\end{pmatrix},
		\end{align*}
		is hyperbolic i.e., $\Hc^0$ has no purely imaginary eigenvalues \cite{pagilla2004controller,aboky2002observers}. 
		Next we derive a condition involving $\alpha,\eta$ for the existence of  a solution to  \eqref{eq-thm3.5}. Consider 
		\begin{align*}
			\det\left(\imath\omega I -\Hc^0\right) &= \det \begin{pmatrix}
				\imath\omega I -A_c^0 & -\alpha^2\Big(I+\frac{BB^{\top}}{\Vert B\Vert^2}\Big)\\
				(1+\eta) I& \imath\omega I+{A_c^0}^{\top}
			\end{pmatrix}.
		\end{align*}
		Shifting first $n-$columns, performing multiplication and using $\Delta(\imath\omega)=\imath\omega I-A_c^0,\; \eta_1=\sqrt{1+\eta}$ we get
		\begin{align*}
			&=(-1)^n\det\left[\Delta^{\top}(\imath\omega)\Delta(\imath\omega)-\alpha^2\eta_1^2\left(I-\frac{BB^{\top}}{\Vert B\Vert^2}\right)\right],\\
			&=(-1)^n\det\left[-\alpha^2\eta_1^2I+\begin{pmatrix}
				\Delta(\imath\omega)\\\alpha\eta_1\frac{B^{\top}}{\Vert B\Vert}
			\end{pmatrix}^{\top}\begin{pmatrix}
				\Delta(\imath\omega)\\\alpha\eta_1\frac{B^{\top}}{\Vert B\Vert}
			\end{pmatrix}\right],
		\end{align*}
		Now from \eqref{eq-3.1.2}, for all $\omega\in \real$,   $\det\left(\imath\omega I -\Hc^0\right)\neq 0$ i.e., $\Hc^0$ cannot have purely imaginary eigenvalues when $\delta_0>\alpha\sqrt{1+\eta}=\alpha\eta_1$.  
	\end{proof}
	Theorem \ref{thm3} shows us that \eqref{eq-thm3.5} may not have  a solution if \eqref{eq-thm3.5a} is violated. In such  situations we perform a coordinate transformation which ensures the condition \eqref{eq-thm3.5a} holds. Now $\delta_0$ can be computed using the algorithm in \cite{Khalil:1173048} as follows.
	\begin{enumerate}
		\item Set $\delta_L =0$, $\delta_U=\Vert M\Vert_2+\Vert N\Vert_2$, and number of iterations $N_d$. Set $j=1$
		\item  If $j>N_d$ exit else go to Step 3.
		\item Compute $\delta_0=\frac{\delta_L+\delta_U}{2}$. Construct $\Hc_{\delta_0}=\begin{pmatrix}
			M &I\\
			N^{\top}N-\delta_0^2 I & -M^{\top}
		\end{pmatrix}$. Check if $\Hc_{\delta_0}$ is hyperbolic. If $\Hc_{\delta_0}$ is hyperbolic $\delta_L = \delta_0$ else $\delta_U=\delta_0$. Set $j=j+1$ and go to Step 2.
	\end{enumerate}
	\subsection{Coordinate Transformation}
	\label{coordinate-transform-1}
	To ensure \eqref{eq-thm3.5a} is true, we have to decrease $\alpha$ and increase $\delta_0$ by using a suitable coordinate transformation $\Tc$. Let
	\begin{align}
		\label{peq-3.2.1}
		\notag	& x = \Tc\phi, \quad \overline{A} = {\Tc}^{-1}A\Tc,\quad \overline{B}={\Tc}^{-1}B,\\
		& \overline{B}_w = {\Tc}^{-1}\quad \overline{C} = C\Tc, \quad \overline{K}=K\Tc.
	\end{align}
	The dynamics now becomes,
	\begin{align*}
		\notag	& \dot{\phi} = \overline{A}\phi+\overline{B}u+\overline{B}_w w,\\
		& z = \overline{C}\phi+Du, \quad u = \overline{K}\phi.
	\end{align*}
	We compute the new Lipschitz constant to be $\overline{\alpha}$ and write the new control objective  using Theorem \ref{thm1}, Theorem \ref{thm2} as,
	\begin{align}
		\label{peq-3.2.3}
		\int_{0}^{\infty}z^{\top}z\;dt \leq\frac{1}{\mu} \tr\left(\overline{B}_w^{\top}\overline{P}\;\overline{B}_w\right),
	\end{align}
	where $\overline{P}=\overline{P}^{\top},\overline{P}\succ 0$ is the solution of 
	\begin{align}
		\label{peq-3.2.4}
		\overline{A}_c^{\top}\overline{P}+\overline{P}\;\overline{A}_c+\overline{\alpha}^2\overline{P}\;\overline{P}+I+\mu \overline{C}_z^{\top}\overline{C}_z=0,
	\end{align}
	where $\overline{A}_c=\overline{A}+\overline{B}\;\overline{K}$ and $\Cb_z=\Cb+D\Kb$. Note that in \eqref{eq-thm1-1} we have $C_z^{\top}C_z$ while in \eqref{peq-3.2.4} we have $\mu \overline{C}_z^{\top}\overline{C}_z$. We explain the  significance of $\mu$ later in the computation procedure. Now using \eqref{peq-3.2.3} and \eqref{peq-3.2.4}, we write \eqref{eq-2.1.2} in the transformed coordinates using $\tr\left(\Bb_w^{\top}\Pb\;\Bb_w\right)=\tr\left(\Pb\;\Bb_w\Bb_w^{\top}\right)$ as,
	\begin{align}\label{peq-3.2.5}
		\underset{d,\overline{K},\overline{P}}{\min} \qquad &
		\fb=\overline{\beta}_df_d\left(d\right)+\overline{\beta}_c\tr\left(\Pb\;\Bb_w\Bb_w^{\top}\right)
		\\
		\notag \text{s.t.}  \qquad
		& 	\overline{A}_c^{\top}\overline{P}+\overline{P}\;\overline{A}_c+\overline{\alpha}^2\overline{P}\;\overline{P}+I+\mu \overline{C}_z^{\top}\overline{C}_z=0,\\
		\notag&\underline{d}\leq d\leq \overline{d},\quad \overline{P}\succ 0.
	\end{align}
	Here $\overline{\beta}_d,\overline{\beta}_c$ are the new optimization weights which can be different from $\beta_d,\beta_c$. Now we solve \eqref{peq-3.2.5} using a  gradient descent method \cite{bertsekas2018nonlinear} to compute optimal $d, \overline{K}$.
	
	To compute the initial stabilizing controller $\Kb^0$ for the transformed system we first construct $\Acb^0=\Tc^{-1}(A_c^0)\Tc=\Tc^{-1}(A+BK_p^0)\Tc$. Using the transformed system $\Acb^0,\Bb,\alphab$ and appropriate $ \etab$ compute  $\deltab_0$ such that \eqref{eq-thm3.5a} is fulfilled. Solve \eqref{eq-thm3.5} to get $\Pb^0$ and  $\Kb_s^0=\frac{-\alphab^2\Bb^{\top}\Pb^0}{2}$. Now for the transformed system we get the initial stabilizing controller gain as $\Kb^0= K_p^0\Tc+\alphab^2\frac{\Kb^0_s}{\Vert \Bb\Vert^2}$. Note that $\eta$ and $\etab$ may be different.
	\subsection{Gradient Computation}
	\label{gradient-compuatation-1}
	In this section we compute the gradient of $\fb$ in \eqref{peq-3.2.5} with respect to the design variable $d$ and the controller gain matrix $\Kb$. For the gradient with respect to $\Kb$ we follow the approach in \cite{rautert1997computational}.
	\begin{lemma}\longthmtitle{Gradient of co-design objective function}
		\label{lem1}
		Consider system \eqref{eq-2.0.1} and co-design problem  in \eqref{peq-3.2.5}. Let $\fb_c=\tr\left(\Pb\;\overline{B}_w^{\top}\overline{B}_w\right)$ and $\Acbb=\Acb+\alphab^2\Pb$ then the following is true.
		\begin{enumerate}
			\item The gradient of the objective function with respect to the $i^{th}$ component $d_i$ of the design variable $d$  is, $\frac{\partial \fb}{\partial d_i} = \overline{\beta}_d\frac{\partial f_d}{\partial d_i}+\overline{\beta}_c\frac{\partial \fb_c}{\partial d_i}$ where $\frac{\partial \fb_c}{\partial d_i}=\tr\left(\frac{\partial \overline{P}}{\partial d_i}\overline{B}_w\overline{B}_w^{\top}\right)$,  $\frac{\partial \overline{P}}{\partial d_i}$ is the solution of 
			\begin{align}
				\label{eq-lem1-1}
				\Acbb^{\top}\frac{\partial \Pb}{\partial d_i}+\frac{\partial \Pb}{\partial d_i}\Acbb+{\frac{\partial\overline{A}_c}{\partial d_i}}^{\top}\overline{P}+\overline{P}\frac{\partial \overline{A}_c}{\partial d_i}=0.	
			\end{align}
			\item The gradient of the objective function with respect to the controller gain variable  $\overline{K}$ is $\nabla_{\overline{K}}\fb=\overline{\beta}_c\nabla_{\overline{K}}\fb_c=2\overline{\beta}_c\left(\overline{B}^{\top}\overline{P}+\mu R\overline{K}\right)\overline{L}$ where $\overline{P}$ is the solution to  \eqref{peq-3.2.4}	and $\overline{L}=\overline{L}^{\top}$ is the solution to 
			\begin{align}
				\label{eq-lem1-2}
				\Lb\;\Acbb^{\top}+\Acbb\;\Lb+\Bb_w\Bb_w^{\top}=0.		
			\end{align}
		\end{enumerate}
		\begin{proof}
			\begin{enumerate}
				\item $f_d$ is a function only of the design variable $d$. To compute $\frac{\partial \fb_c}{\partial d_i}$, the quantity $\frac{\partial \overline{P}}{\partial d_i}$ is obtained  by   partially differentiating \eqref{peq-3.2.5} with respect to $d_i$ to give \eqref{eq-lem1-1}. 
				\item  Differentiating the constraint equation in \eqref{peq-3.2.5} with respect to $\overline{K}$ and rearranging gives,
				\begin{align}
					\label{eq-lem1-3}
					\notag&\Acb^{\top}\Pb_{\Kb}d\Kb+\Pb_{\Kb}d\Kb\;\Acb+\left(\Bb d\Kb \right)^{\top}\Pb+\Pb\left(\Bb d\Kb \right)+\alphab^2\Pb_{\Kb}d\Kb\;\Pb\\
					\notag&\hspace{0.3cm}+\alphab^2\Pb\;\Pb_{\Kb}d\Kb+\mu\left(Dd\Kb \right)^{\top}\Cb_z+\mu\Cb_z^{\top}\left(Dd\Kb\right)=0,\\
					& \Acbb^{\top}\Pb_{\Kb}d\Kb+\Pb_{\Kb}d\Kb\;\Acbb+\left(\Bb d\Kb \right)^{\top}\Pb+\Pb\left(\Bb d\Kb \right)\\
					\notag &\hspace{2cm}+\mu\left(Dd\Kb \right)^{\top}\Cb_z+\mu\Cb_z^{\top}\left(Dd\Kb\right)=0,
				\end{align}
				where $\Pb_{\Kb}$ is the derivative of $\Pb$ with respect to $\Kb$ and $d\Kb$ is the differential of $\Kb$. Pre-multiplying \eqref{eq-lem1-3} by $\Lb$,  post-multiply \eqref{eq-lem1-2} by $\Pb_{\Kb}d\Kb$ and taking trace gives,
				\begin{subequations}\label{eq-lem1-4}
					\begin{align}
						\notag&\tr(\Lb\;\Acbb^{\top}\Pb_{\Kb}d\Kb+\Lb\;\Pb_{\Kb}d\Kb\;\Acbb)\\
						\label{eq-lem1-4b}&\hspace{2cm}=-2\tr\big((\Lb\;\Pb\;\Bb+\mu\Lb\;\Cb_z^{\top}D)d\Kb\big),\\
						\label{eq-lem1-4c}&\hspace{2cm}=-\tr\left(\Bb_w\Bb_w^{\top}\Pb_{\Kb}d\Kb \right).
					\end{align}
				\end{subequations}
				Now,
				\begin{align}
					\label{eq-lem1-5}
					d\fb_c = \tr\left(\nabla_{\Kb}f_c^{\top}dF\right)=\tr\left(\Bb_w\Bb_w^{\top}\Pb_{\Kb}d\Kb\right).
				\end{align}
				From \eqref{eq-lem1-4b}, \eqref{eq-lem1-4c}, \eqref{eq-lem1-5} and assumption ({\bfseries A3}) i.e., $\Cb_z^{\top}D = \Kb^{\top}D^{\top}D=\Kb^{\top}R$ we get,
				\begin{align*}
					\nabla_{\Kb}\fb=\nabla_{\Kb}\fb_c=2\betab_c\left(\Bb^{\top}\Pb+\mu R\Kb\right)\Lb.
				\end{align*}
			\end{enumerate}
		\end{proof}
	\end{lemma}
	In the gradient based co-design algorithm we may need to solve \eqref{peq-3.2.4} repeatedly while computing the gradient at each iteration. For $\mu=1$  the  Hamiltonian matrix of \eqref{peq-3.2.4} may not  be hyperbolic leading to a breakdown of the co-design algorithm. Hence, an appropriate selection of $\mu$ ensures  a solution to \eqref{peq-3.2.4} at each iteration for smooth running of the co-design algorithm.
	\subsection{Co-design Algorithm}
	\label{codesign-algo1}
	In this section we present the iterative co-design algorithm to compute the optimized design and control variables $d^{opt},\Kb^{opt}$ in a step-wise format as follows.
	\begin{enumerate}
		\item Set design bounds, $\underline{d},\overline{d}$, initial starting design $d^0$, Lipschitz constant $\alpha$, $B$ and construct $A$. 
		\item If $A$ is Hurwitz  then $K^0_p=0$. If $A$ is not Hurwitz then compute $K^0_p$ required in Theorem \ref{thm3}. Note that we can use pole placement or LQR or any other other heuristic/deterministic method to compute $K_p^0$ \cite{skogestad2005multivariable}. 
		\item Select appropriate $\eta$ and compute $\delta_0$ in Theorem \ref{thm3}.  We have two cases here as follows.
		\begin{enumerate}
			\item If \eqref{eq-thm3.5a} is fulfilled then set $\Tc = I,\;\etab=\eta$,  proceed to Step 4.
			\item If \eqref{eq-thm3.5a} is not fulfilled then choose appropriate $\Tc,\etab$ which fulfill \eqref{eq-thm3.5a} and proceed to Step 4. Note that here $\etab$ may or may not be  equal to $\eta$. 
		\end{enumerate}
		\item Using $\Tc$, perform coordinate transformation as per Section \ref{coordinate-transform-1} and compute
		$\Kb^0$.
		\item Choose $\mu$ such that the Hamiltonian of \eqref{peq-3.2.4} for the transformed system is hyperbolic \cite{aboky2002observers,pagilla2004controller}. Set $j=0$, $\Kb^j=\Kb^0,d^j = d^0$, tolerance $\varepsilon_g$, weights $\betab_d,\betab_c$. 
		\item Compute gradients $\nabla_d\fb(d^j,\Kb^j)$ and $\nabla_{\Kb}\fb(d^j,\Kb^j)$ using Lemma \ref{lem1}.
		\item Use Armijo rule \cite[Section 1.2]{bertsekas2018nonlinear} stated next to find the step-size $s^j$. Set step-size $s^j=1$, $0<\nu,\zeta<1$. Repeat $s^j=\nu s^j$ until $\fb(d^j-s^j\nabla_d \fb,\Kb^j-s^j\nabla_{\Kb}\fb)<\fb(d^j,\Kb^j)-s^j\zeta(\nabla_d^{\top}\fb\nabla_d\fb+\tr(\nabla_{\Kb}^{\top}\fb\nabla_{\Kb}\fb))$. Note that typically $\nu=0.5,\zeta=0.3$.
		\item Compute $d^{j+1}=d^j-s^j\nabla_d \fb$, $\Kb^{j+1}=\Kb^j-s^j\nabla_{\Kb}\fb$. If $\Vert d^{j+1}-d^j\Vert+\Vert \Kb^{j+1}-\Kb^j\Vert\leq \varepsilon_g$ then $d^{opt} = d^j,\Kb^{opt}=\Kb^j$ and exit else $d^j = d^{j+1},\;\Kb^j=\Kb^{j+1}$ and go to Step 6.
	\end{enumerate} 
	Our next result  shows how we can synthesize stabilizing controller $K^{opt}$ for system \eqref{eq-2.0.1} from $\Kb^{opt}$.  
	\begin{theorem}\longthmtitle{Stabilizing controller gain for original system}
		\label{thm4}
		Consider system \eqref{eq-2.0.1} and its transformation \eqref{peq-3.2.1} done using the non-singular transformation matrix $\Tc$. If $\Kb^{opt}$ is the stabilizing controller with the design $d^{opt}$ obtained from the co-design algorithm for the transformed system \eqref{peq-3.2.1} then $K^{opt}=\Kb^{opt}\Tc^{-1}$ is the stabilizing controller for the original system \eqref{eq-2.0.1}. 
	\end{theorem}
	\begin{proof}
		From Section \ref{coordinate-transform-1} we have $\Ab(d^{opt})+\Bb\;\Kb^{opt} = \Tc^{-1}(A(d^{opt})+BK^{opt})\Tc$. As $\Tc$ is a similarity transformation $(\Ab(d^{opt})+\Bb\;\Kb^{opt})$ and $A(d^{opt})+BK^{opt}$ have the same eigenvalues \cite{horn2012matrix}. As $(\Ab(d^{opt})+\Bb\;\Kb^{opt})$ is Hurwitz  so $(A(d^{opt})+BK^{opt})$ is also Hurwitz. With $\Kb^{opt}$,  $z\rightarrow 0$ as $t\rightarrow \infty$ for the transformed system. But  $x=\Tc z \rightarrow 0$ as $t\rightarrow \infty$, so $K^{opt}$ is a stabilizing controller for the original system.
	\end{proof}
	Next we present an example illustrating our co-design procedure.
	\section{Co-Design of Single-Link Manipulator}
	\label{example-1}
	Consider a single-link flexible manipulator setup consisting of a single flexible link driven by a DC motor. The dynamics is given by the following system parameter matrices \cite{raghavan1994observer},
	\begin{align}
		\label{peq-4.1.1}
		&x = \begin{pmatrix}
			x_1 \\ x_2 \\ x_3 \\ x_4
		\end{pmatrix},
		A = \begin{pmatrix}
			0 & 1 & 0 & 0\\
			\frac{-k_m}{J_m} &\frac{-b_m}{J_m} & \frac{k_m}{J_m} & 0\\
			0 & 0 &1 & 0\\
			\frac{k_m}{J_l} & 0& \frac{-k_m}{J_l} & 0
		\end{pmatrix},\Phi=\begin{pmatrix}
			0 \\ 0 \\ 0 \\ \frac{-mgl}{J_l}\sin x_3
		\end{pmatrix},\\
		\notag&B = \begin{pmatrix}
			0\\ \frac{k_t}{J_m}\\0\\0
		\end{pmatrix},B_w = \begin{pmatrix}
			0\\ 0 \\ 1 \\0
		\end{pmatrix},C = \begin{pmatrix}
			1 & 0 & 0 & 0\\
			0 & 1 & 0 & 0\\
			0 & 0 & 1 & 0\\
			0 & 0 & 0 & 0 
		\end{pmatrix},
		D = \begin{pmatrix}
			0 \\ 0 \\0 \\1
		\end{pmatrix}.
	\end{align}
	Here $x_1$ and $x_3$ are the angular rotations of  the motor and the link respectively, and $x_2$ and $x_4$ are their  angular velocities. $J_m$ is  inertia of the motor, $J_l$ is inertia of the link, $k_m$ is the torsional constant, $b_m$ is the viscous damping constant, $k_t$ is the amplifier gain, $m$ is  mass of the link, $l$ is distance to the center of mass of the link and $g$ is the gravitational acceleration. The parameter values are as follows:
	\begin{align*}
		&J_m = 0.0037;\;J_l = 0.0093;\;m=0.021;\;l=0.15\;\\
		&k_m = 0.18;\; k_t=0.08;\;g=9.81.
	\end{align*}
	We take damping  as the design variable $d=b_m$ with the bounds $0.002\leq b_m \leq 0.1$.  
	The Lipschitz constant $\alpha = \frac{mgl}{J_l}=3.33$. Using  $d^0=0.0046$ we construct $A$ whose initial poles are at $-0.44\pm8.21,\;-0.36,\; 0$. Following the co-design algorithm in Section \ref{codesign-algo1}, as $A$ is not Hurwitz we first compute $K^0_p$ by pole placement. We select the poles of $A_c^0=A+BK_p^0$ as $-9,-7,-6,-4$ and compute $K_p^0=\begin{pmatrix}
		-8.27 &   -1.15 &    4.67 &   -1.22
	\end{pmatrix}$. Now we select $\eta = 0.0001$ and compute $\delta_0 = 0.86$ as per Section \ref{initial-gain-1}. But $\alpha\sqrt{1+\eta}=3.33$ which gives $\delta_0<\alpha\sqrt{1+\eta}$ violating Theorem \ref{thm3}. We select $\Tc = \diag(1,\;1,\; 1,\; 10)$ and compute the transformed system as per Section \ref{coordinate-transform-1}. For the transformed system   $\alphab=0.333,\;\etab=\eta=0.0001$ we get $\deltab_0=0.47$, $\alphab\sqrt{1+\etab}=0.333$ and $\deltab_0>\alphab\sqrt{1+\etab}$. The initial stabilizing gain is $\Kb^0=\begin{pmatrix}
		-8.28 &   -1.15 &    4.67 &  -12.21
	\end{pmatrix}$ computed as explained in Section \ref{coordinate-transform-1}. We set $\mu=0.01,\;\varepsilon_g=10^{-3}$ and perform gradient descent procedure. We get the optimal design and control variables as $d^{opt} = 0.06,\;\Kb^{opt}=\begin{pmatrix}
		-9.37&   -1.32 &    4.46 & -11.45
	\end{pmatrix}$ and $ K^{opt}=\begin{pmatrix}
		-9.37&   -1.32 &    4.46 & -1.15
	\end{pmatrix}$. The improvement in performance of the transformed system is $\frac{\fb(d^0,\Kb^0)-\fb(d^{opt},\Kb^{opt})}{\fb(d^0,\Kb^0)}\times 100=34\%$. Note that we  measure performance improvement for the transformed system as for the original system \eqref{eq-thm1-1} has no solution.
	\begin{figure}[htp]
		\centering
		\hspace*{-1em}
		\subfloat[Evolution of $x_1$]{\label{figx1}\includegraphics[width=\columnwidth,height =3cm]{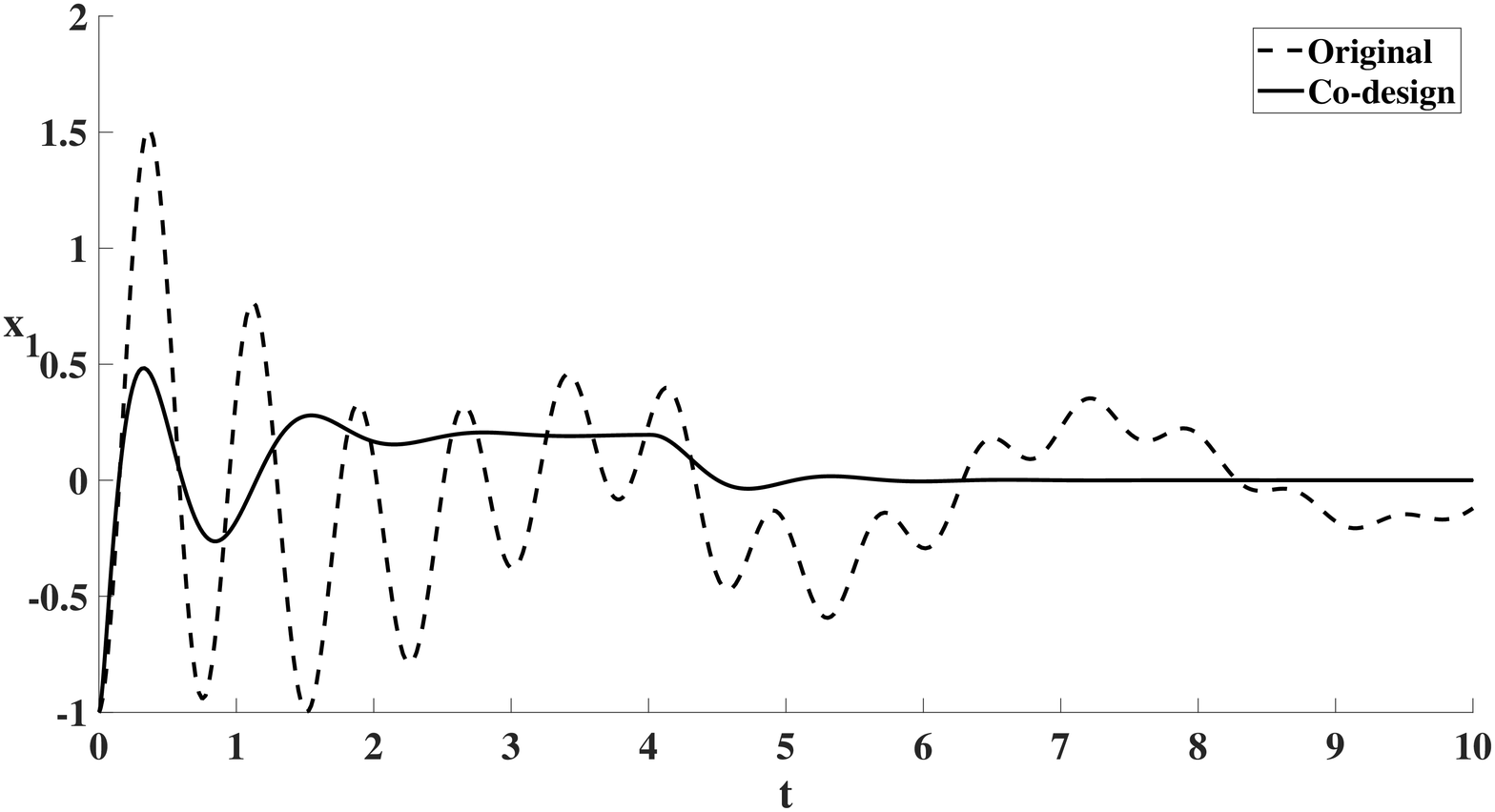} }\\
		\hspace*{-1em}
		\subfloat[Evolution of $x_2$]{\label{figx2}\includegraphics[width=\columnwidth,height =3cm]{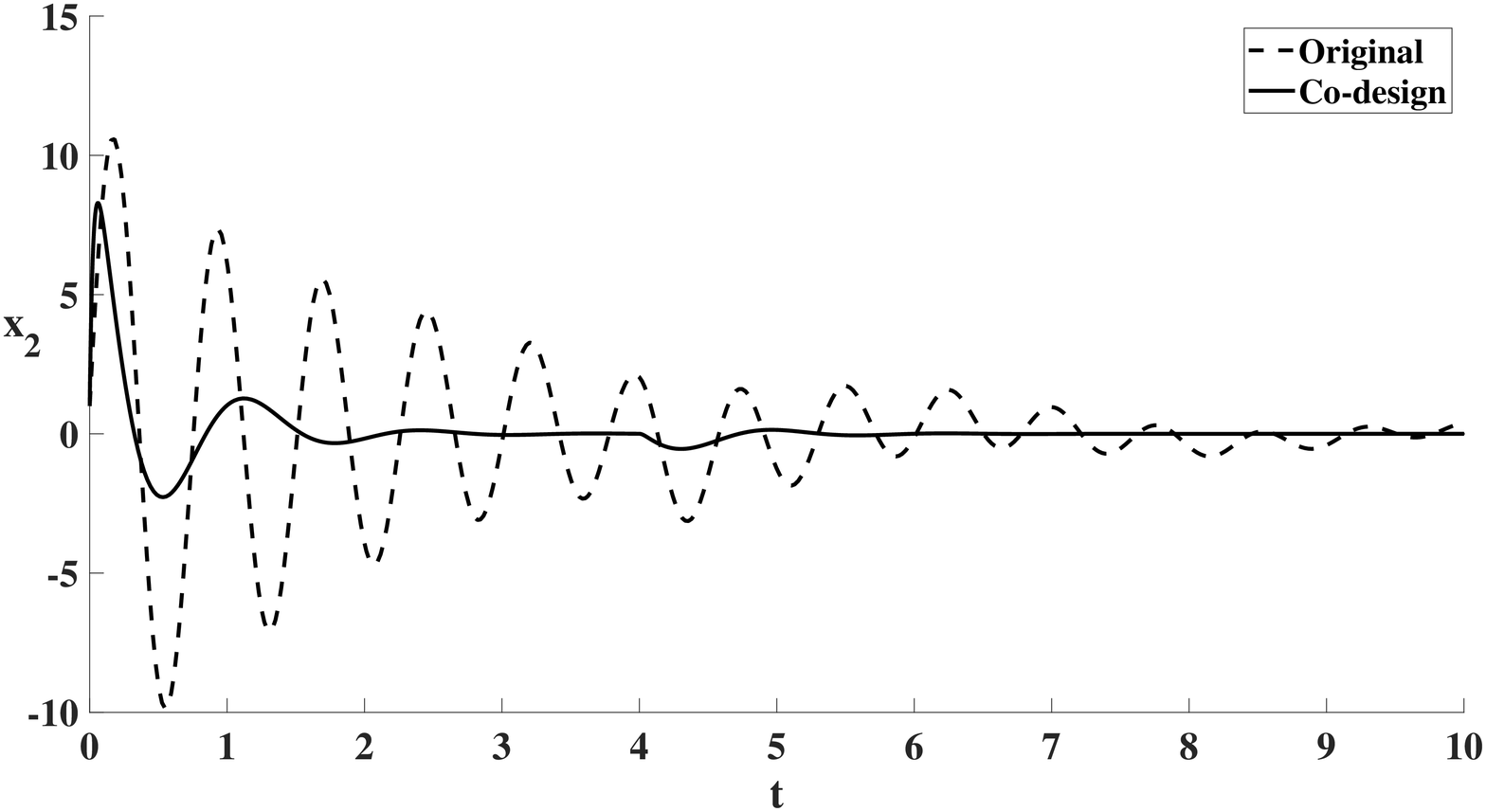} }\\
		\hspace*{-1em}
		\subfloat[Evolution of $x_3$]{\label{figx3}\includegraphics[width=\columnwidth,height =3cm]{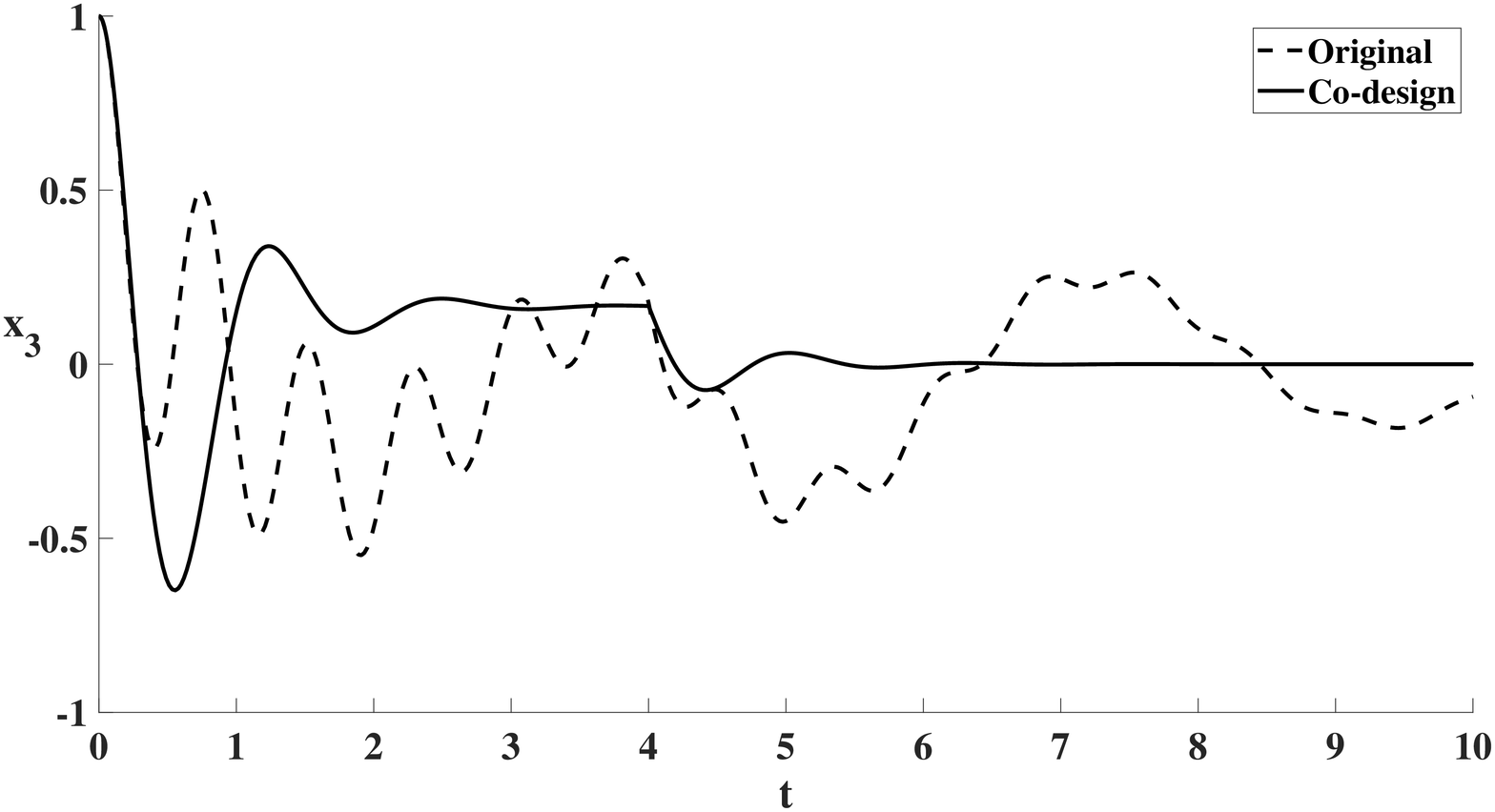} }
		\\
		\hspace*{-1em}
		\subfloat[Evolution of $x_4$]{\label{figx4}\includegraphics[width=\columnwidth,height =3cm]{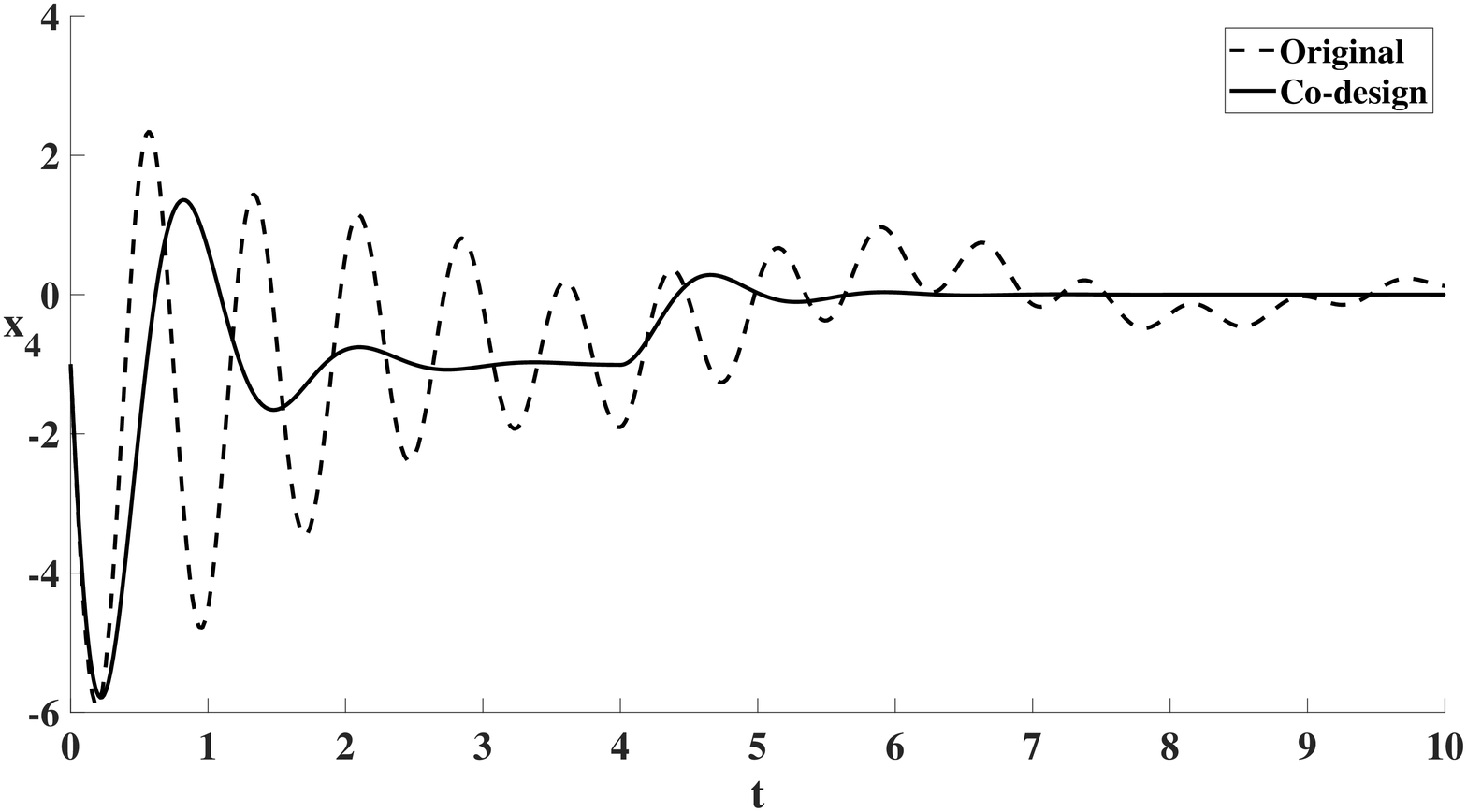} }
		\caption{Evolution of states of the system with time. }%
		\label{fig:evolution}%
	\end{figure} 
	We simulate the system  for $w=1$ applied for the time span  $0\leq t\leq4$  with initial condition $x_0 = \begin{pmatrix}
		-1 & 1 & 1 & -1
	\end{pmatrix}^{\top}$ and show the result in Figure \ref{fig:evolution}.  We observe that the co-designed system shows an improvement in  performance.
	\section{Conclusion}
	\label{conclusion-1}
	In this letter we have studied the co-design optimization problem for systems with Lipschitz nonlinear dynamics. We propose a novel time independent reformulation of the  co-design problem with a quadratic matrix equation as constraint ensuring system stability. We then propose a gradient based iterative method to compute a solution of the co-design problem. Our future work includes providing convergence and optimality guarantees to the proposed co-design solution method and extending it to general nonlinear systems. 
	
	\bibliographystyle{IEEEtran}
	\bibliography{ref1}

\end{document}